\documentclass{aastex}

\def\app{\ifmmode{\sim}\else${\sim\,}$\fi}
\def\deg{\ifmmode^\circ\,\else$^\circ\,$\fi}

\newcommand{\Lya}{H $Ly_{\alpha}$\,}
\newcommand{\Lyb}{H $Ly_{\beta}$\,}
\newcommand{\Lyg}{H $Ly_{\gamma}$\,}

\newcommand{\ten}[1] {\ifmmode {10^{#1}} \else {$ 10^{#1} $} \fi}
\newcommand{\pwr}[2] {\ifmmode {#1 \;x\; 10^{#2}} \else {$ #1 \;x\; 10^{#2} $} \fi}

\begin{document}

\title{ 
Reanalyses of {\em Voyager} Ultraviolet Spectrometer Limits to the
EUV and FUV Diffuse Astronomical Flux 
}

\author{Jerry Edelstein}
\email{jerrye@ssl.berkeley.edu}
\author{Stuart Bowyer}
\email{bowyer@ssl.berkeley.edu}
\author{Michael Lampton}
\email{mlampton@ssl.berkeley.edu}
\affil{Space Sciences Laboratory, University of California,
Berkeley, CA 94720}

\begin{abstract}

We re-examine {\em Voyager}
Ultraviolet Spectrograph data used
to establish upper limits to the $\lambda$ 500--900 \AA\
and $\lambda$ 900--1100 \AA\ cosmic diffuse background.
The measurement of diffuse flux with the {\em Voyager} UVS data
requires complex corrections for noise sources
which are far larger than the astronomical signal.
In the analyses carried out to date,
the upper limits obtained on the diffuse background
show statistical anomalies which indicate that
substantial systematic errors are present.
We detail these anomalies and identify specific
problems with the analysis.
We derive statistically robust 
2~$\sigma$ upper limits 
for continuum flux of
570~photons~s$^{-1}$~cm$^{-2}$~ster$^{-1}$~\AA$^{-1}$
and 
for the 1000 \AA~ diffuse line flux
of 11,790~photons~s$^{-1}$~cm$^{-2}$~ster$^{-1}$.
The true limits may be substantially higher
because of unknown systematic uncertainties.
The new statistical limits alone are insufficient
to support previous conclusions 
based on the {\em Voyager} data including work on 
the character of interstellar dust
and estimates of the diffuse extragalactic far UV background
as absorbed by intergalactic dust.

\end{abstract}

\label{voyager}

\section{Introduction}

Measurements of the far ultraviolet (FUV: $\lambda\lambda$ 900--1500 \AA) 
and extreme ultraviolet (EUV: $\lambda\lambda$ 90-900 \AA)
diffuse background are of considerable interest.
The FUV background longer than 1200 \AA\ has been
categorized in general,
(see Bowyer 1991 for a review)
but minimal information exists on the band from $\lambda$ 900--1200 \AA.
Only upper limits to the general EUV cosmic background exist
(see Bowyer et al 1996 for a review of the EUV background);
the lowest limits for the EUV background from 500--900 \AA\ 
have been reported by 
Edelstein et al (1999).
Possible contributors to the  $\lambda$ 900--1200 \AA~ 
background include
radiation from the hot component of the interstellar medium (ISM),
starlight scattered from interstellar dust, 
emission from fluorescing molecular hydrogen,
resonantly scattered radiation from inflowing
neutral components of the warm ISM,
resonantly scattered geophysical lines, and 
flux from the electromagnetic decay of elementary
particles created in the early universe. 

Holberg (1986) and Murthy et al (1991, 1999) have reported the
lowest limits for the FUV 900--1100 \AA\ background.
These results were obtained using data obtained with the
Ultraviolet Spectrometer on the {\em Voyager}
Spacecraft (Broadfoot et al, 1977).
Because of a variety of instrumental and environmental effects,
the recorded count rate from the {\em Voyager} instrument was substantially
greater than any astronomical diffuse flux;
it is about 40 times larger than the claimed upper limit.
In principle, a sufficiently precise determination
of a stable background will allow an arbitrarily small signal
embedded in this background to be identified.
However, for data with very low signal to noise
even small systematic errors can overwhelm any
possibility of deriving a valid signal.
We have identified several difficulties in the data set and/or 
the analysis of Holberg (1986) 
and Murthy et al (1991) which suggest
that significant systematic errors (as compared to the upper
limits obtained) are affecting the results.
These difficulties are also relevant to the results of Murthy et al (1999) 
because they incorporate the data reduction procedures of Holberg (1986).
In this paper we discuss these problems, 
and identify some of the systematic effects
which are present.

\section{Observations and Analysis}

Holberg (1986) used 1,508,198 sec of observation
from four separate (but astronomically equivalent) regions to
obtain limits on a diffuse astronomical flux
over the band from 500--1100 \AA~
of 100 to 400 photons~s$^{-1}$~cm$^{-2}$~ster$^{-1}$~\AA$^{-1}$.
Holberg did not conduct a formal statistical error analyses
and chose an ad-hoc (albeit reasonably considered)
error level to establish his limits.
Murthy et al (1991) used observations from four different regions
of the sky to obtain 2-$\sigma$ upper limits 
on the diffuse flux from 900--1100 \AA~ ranging from 
-38 to 500 photons~s$^{-1}$~cm$^{-2}$~ster$^{-1}$~\AA$^{-1}$
(see Table 1).
Three of these four observations had integrations of $\sim$230,000 sec,
the fourth provided $\sim$83,000 sec of data.
The reduction of both data sets was similar
for the majority of the analysis.
The major difference was that
Murthy et al considered the statistical contribution from only 
one of the many possible contributors to the ultimate error
and employed an additional assumption regarding the statistical 
independence of data sets that is discussed hereafter.

An immediate indication that large systematic errors may be
present in the analysis by Murthy et al (1991) is provided by the data
shown in Table 2 of that reference.
Systematics can easily vitiate statistical analyses,
and the fact that the largest derived fluxes are negative indicates 
that systematic oversubtraction may have taken place.
In that table the authors provide their best estimates
for the background in four view directions.
To assess the likelihood that the four quoted data are free of
systematic error, we have computed 
the probability of each measurement being consistent with zero flux
with a normal probability function.
The results are shown in Table 1.
Under the null hypothesis of zero continuum flux and no systematic
errors being present, the background-subtracted fluxes should
group around zero mean and each would exhibit a reasonable ranking
against a normal distribution, say 0.1 to 0.9.  A negative two $\sigma$
deviation is a rare event with a probability of $P=0.023$.  Finding 
two such events in four independent measurements is much rarer still.
The probability of such an ensemble can be found by combining
all permutations of two special events among four bins, and multiplying
by $P^{2}$ which is the a priori probability of two special events occurring.
This estimate, 6$P^{2}$ = 0.004, reveals the improbability of the 
ensemble being governed by the stated random errors.

A different kind of test yields the same conclusion.  The chi-square
test is an aggregate measure of the total discrepancy represented
by an ensemble of measurements.  As such, it is tolerant of some data
being highly discrepant when other data in the ensemble are less 
discrepant than average.  The total chi square 
statistic for the four reported measurements is 9.6, a value 
reached by bias-free statistics in fewer than 5\% of all ensembles.
This test rejects the hypothesis that the data are fairly described
by their random errors at the 95\% confidence level.  

We note that
if we simply ignore the Murthy et al quoted errors, 
and consider only the scatter in the fluxes assuming
that they are measurement samples of zero flux
and that no systematic errors are present, we can estimate
their uncertainty from their scatter.  In this case the most probable
value of their standard deviation is 211 
photons~s$^{-1}$~cm$^{-2}$~ster$^{-1}$~\AA$^{-1}$, 
a value that would bring the chi squared of their fit to 4.00 
for four sample points.  This estimate then gives a statistical 
2 $\sigma$ upper limit margin of 420 
photons~s$^{-1}$~cm$^{-2}$~ster$^{-1}$~\AA$^{-1}$ for these data.
It is valid to draw statistical conclusions from the set of 
these four data points -- if the errors are not systematic
and are estimated properly.
Indeed Lampton (1994) has shown that robust statistical conclusions 
may be determined
from just two Poisson samples, one for combined signal and background 
and the other for background alone.
However, our major concern with these data is the very great danger of
systematic effects dominating random effects. 
The finding that two independent tests each show that the 
published data set is improbable indicates that substantial 
systematic errors are likely to be present.   

We have re-analyzed the long observation of Holberg (1986) toward
the north Galactic pole. We consider the uncertainties for
each stage of the data reduction process. 
In Table 2 we list the count rate, the respective photon count 
given the integration time, and the uncertainty for each 
of the signal and noise components within a single 
instrumental bin located at 1000 \AA~ of spectral width of $\app$ 10 \AA~ .
These values are typical of the entire data set from 
650 -- 1000 \AA~ although larger 
airglow line contributions exist about the
Ly--$\gamma$ and Ly--$\beta$ spectral regions.
We note that the {\em Voyager} data system registered
three counts for each valid photon event (Murthy et al, 1991);
the numbers for photon events in our Table 2 are 
one-third the product of the count rate and the total integration time.

The principal noise components in the {\em Voyager} data are
background counts produced by the on-board radioactive
power generator, and instrumental scattering of the
interplanetary solar resonance lines of hydrogen and helium.
An estimate of the radioactively induced ``dark-count" background 
was obtained by observing a shadowed area of the spacecraft.
This spectrum was scaled to fit the 650--900 \AA\ region where 
no detectable line emission was anticipated, 
given the instrument's sensitivity and interstellar absorption.
The dark-count background correction
corresponds to 11,000 photon events during the observation and
represents 58\% of the recorded signal.
While this background spectrum could, in principle, be stable,
Holberg (1986) 
shows two very long integrations of this background spectrum in which 
approximately half of the more than 100 spectral channels 
show variations from spectra to spectra of \app 2\%.
These variations are about a factor of two larger
than the 1 $\sigma$ statistical counting variation per channel.
While we consider 2\% of the background count level as an appropriate 
estimate for the uncertainty in the background, 
we simply adopt the statistical count rate uncertainty as the minimum
background uncertainty for our analysis.

The next manipulation of the data is to remove flux from
the strong interplanetary solar resonance lines of \Lya, \Lyb, and 
He  $Ly_{\alpha}$\, which are
instrumentally scattered throughout the recorded spectra.
The de-scattering correction 
represents 7,000 photon events during the observation or
37\% of the recorded signal.
A ground calibration de-scattering matrix was used
that redistributes photons among the spectral bins.
Errors in either the input spectrum or the de-scattering matrix operator
will combine to affect the result.
Any background counts that have not been removed from the spectrum
will be redistributed according the matrix operator.
Errors within the de-scattering matrix and its interaction with
possible errors in background removal are difficult to assess
but must be present at some level 
as demonstrated by the application of the de-scattering technique to 
an observation of a hot stellar continuum which 
removed only 93\% of the light scattered from above the Lyman limit
from the 600--900 \AA\ band stellar flux
(Holberg and Barber 1985).
We conclude that uncertainties in the de-scattering process
are likely to introduce errors
that will add to the statistical counting uncertainty.
These additional errors cannot be determined from the data presented
so we adopt the counting error as the minimum uncertainty for this component.

After the de-scattering process, 
modeled interplanetary solar resonance lines are subtracted 
from the spectrum to obtain a residual spectrum
from which the limits to an astronomical flux are derived.
The uncertainty in the subtraction of interplanetary lines
varies across the band, but, excluding the region about Ly--$\beta$,
the uncertainty due to line flux removal is not a dominant factor.  
The line modeling correction at 1000 \AA\
represents 500 photon events during the observation
or 2.6\% of the recorded signal.
Systematic error must be present 
in the line modeling subtraction at some level.
For example, Holberg (1986) notes that
incomplete removal of the \Lya line renders these results
invalid for $\lambda > 1100$ \AA.
The magnitude of this uncertainty cannot be determined
from the data presented,
but must be an addition to the statistical uncertainties.

The residual flux is determined 
by subtracting the background components from the recorded signal
and corresponds to 500 photon events during the observation,
or 2.6\% of the recorded signal.
At a minimum, the uncertainty in the residual flux 
is dependent on the errors in each of the signal and background
components summed in quadrature. 
The minimum combined uncertainty, 
considering our discussion above,
is 194 photon events (see Table 2).  
Consequently, the 2 $\sigma$ upper limit to the photon events
occurring within a single channel during the observation, 
which equals the value of the residual net signal 
plus twice the standard error, is 888 photon events.
We can convert this single channel value to physical continuum units,
or 570 photons~s$^{-1}$~cm$^{-2}$~ster$^{-1}$~\AA$^{-1}$,
given the observation time of $1.5 \times 10^6$~sec and
a flux conversion factor determined from the calibrated Voyager spectrum of 
the white dwarf PG1034+001 of 
$3.21 \times 10^5 $photons~s$^{-1}$~cm$^{-2}$~ster$^{-1}$~\AA$^{-1}$
per UVS count per second (Holberg private communication 1997).

To determine the 2 $\sigma$ upper limit to a diffuse monochromatic 
emission line occurring in the spectra we consider that the line profile 
characteristic width is three channels wide, or $\app$ 30 \AA\ .
If we take the favorable view that errors in 
three adjacent channels are statistically independent (but see below) 
then the minimum combined uncertainty in a 
channel is 194 $\times \sqrt{n}$ photon events and the 2 $\sigma$ upper limit 
is 612 photon events per single channel.  Using the flux conversion
factor and the 30 \AA\ characteristic line width we determine 
a physical value of the minimum 2 $\sigma$ upper limit to 
single line emission intensity of 
11,790 photons~s$^{-1}$~cm$^{-2}$~ster$^{-1}$.

We note that the Murthy et al (1991 \& 1999) 
derivation of continuum limits using the 
900--1100 \AA\ band {\em Voyager} data 
followed the same general procedures
as Holberg (1986) and consequently the points
raised above are also valid for their analysis.
In addition these authors assumed that each spectral 
channel's errors were uncorrelated.  However, the de-scattering process 
explicitly takes counts from each channel and redistributes
them to other channels, which directly violates the assumption
of statistical independence.
Hence this approach is not valid $prima facia$ and its use
cannot be justified. 

\section{Conclusions}

Using the published {\em Voyager} data of Holberg (1986) 
for observations of the north Galactic pole
we estimate a 2~$\sigma$ upper limit for 
the 1000 \AA~ diffuse line flux 
of 11,790~photons~s$^{-1}$~cm$^{-2}$~ster$^{-1}$.
The limit to continuum flux is 
570~photons~s$^{-1}$~cm$^{-2}$~ster$^{-1}$~\AA$^{-1}$.
These statistically robust 2~$\sigma$ upper limits
are in contrast to the published limits of 
100 -- 200 ~photons~s$^{-1}$~cm$^{-2}$~ster$^{-1}$~\AA$^{-1}$
as a 1 $\sigma$ limit (Murthy et al 1991)
and 
6,000 ~photons~s$^{-1}$~cm$^{-2}$~ster$^{-1}$,
with no $\sigma$ level quoted (Holberg 1986).
These new statistical results do not 
allow for probable systematic effects discussed 
in the text that are likely raise these limits 
by some additional unknown amount.
Continuum limits based on the new statistical limits alone
are greater than the values needed
to support a variety of conclusions of Murthy et al (1991)
including the derived properties of interstellar dust.
We also note that Overduin and Wessin (1997) and Overduin et al (1999)
used the Murthey etal (1991) dust results to derive the extragalactic 
far UV background after absorption by intergalactic dust.
These results may now be compromised.

{\bf Acknowledgements}
We acknowledge interesting discussions with Richard Henry.

\clearpage
\begin{table}[hbtp]
\caption{Probability of the Reported Fluxes Being Consistent with Zero}
\begin{center}
\begin{tabular}{|lccc|}\hline
 & Reported Flux & Probability & 2 $\sigma$ Upper Limit \\[0.1in]
 & CU & & CU \\ \hline \hline  
Target A & -264$\pm$132 & 0.023   & 0\\
Target B & 48$\pm$227 & 0.6 & 502 \\
Target C & -72$\pm$138 & 0.28 & 200 \\
Target D & -320$\pm$141 & 0.013 & -38 \\ \hline 
 & Total Ensemble Permutation probability:  & 0.004 & \\
 & Total Ensemble Chi-square probability:  & $<$0.05 & \\ \hline
\end{tabular}
\end{center}
Note:  CU $=$ photons~s$^{-1}$~cm$^{-2}$~ster$^{-1}$~\AA$^{-1}$
\end{table}

\clearpage
\begin{table}[hbt]
\caption{{\em Voyager} Signal Components per channel}
\begin{center}
\begin{tabular}{|l|c|c|c|} \hline 
Component &Counting & Photon & Minimum 			\\
	  &Rate	    & Events & Uncertainty (1 $\sigma$)  \\ \hline\hline
{\bf Recorded Signal}		& 0.038 & 19,000 & 138 \\[0.1in]
{\bf Background Components}	& & &   \\
\hspace{0.2in} Dark-Count & 0.022 & 11,000 & 105 \\
\hspace{0.2in} Instrumental Scattering	&       &             &\\
\hspace{0.3in}  of Geocoronal Lines   & 0.014 & 7,000  & 84 \\
\hspace{0.2in} Interplanetary \Lyb \& \Lyg & 0.001 & 500  & 22 \\ \hline\hline
Signal minus Background	& 0.001  & 500 	& 	  	\\
Combined Uncertainty	& 	 & 	& 194 \\ \hline \hline
2 $\sigma$ Upper Limit  & 	 & 	& 888 \\ \hline \hline
\end{tabular}
\end{center}
\end{table}

\end{document}